\documentclass{easychair}

\pdfoutput=1

\newcommand{\xs}[1]{Section~\ref{#1}}

\newcommand{\xe}[1]{Equation~\ref{#1}}

\newcommand{\gipsy}{{GIPSY\index{GIPSY}}}

\newcommand{\lucid}{{Lucid\index{Lucid}}}

\newcommand{\flucid}{{Forensic Lucid\index{Forensic Lucid}}}

\newcommand{\trans}{$\psi$}

\newcommand{\invtrans}{$\Psi^{-1}$}

\newcommand{\dmarf}[0]{DMARF\index{MARF!Distributed}\index{Frameworks!Distributed MARF}\index{Libraries!Distributed MARF}}

\newcommand{\lucidL}[1]{{$\mathit{Lucid}$}($L$) }

		{}

\def\myvert{\raise 2.27pt \hbox{\vrule depth 0pt height 8pt width 0.2mm}}
\def\myarrow{\hspace*{0.43mm}%
             \raise 2.29pt\hbox{\vrule depth 0pt height 8pt width 0.16mm}%
             \hspace*{-0.32mm}%
             $\longrightarrow$
             \ %
             }

\newcommand{\assl}{ASSL\index{ASSL}}
\newcommand{\smart}{S.M.A.R.T.}

\begin{document}

%\title{The Role of Self-Forensics in Crash Investigations and Event Reconstruction}
\title{The Role of Self-Forensics in Vehicle Crash Investigations and Event Reconstruction}
\titlerunning{Self-Forensics in Vehicle Crash Investigations}

\author{Serguei A. Mokhov\\
Department of Computer Science and Software Engineering\\
Concordia University, Montreal, Canada\\
\url{mokhov@cse.concordia.ca}\\
}

\authorrunning{Mokhov}

\maketitle

\begin{abstract}
This paper further introduces and formalizes a novel concept of
self-forensics for automotive vehicles, specified in the {\flucid}
language. We argue that self-forensics, with the forensics taken
out of the cybercrime domain, is applicable to ``self-dissection'' of
intelligent vehicles and hardware systems for automated incident and
anomaly analysis and event reconstruction by the software with or
without the aid of the engineering teams in a variety of forensic
scenarios. We propose a formal design, requirements, and specification
of the self-forensic enabled units (similar to blackboxes) in vehicles
that will help investigation of incidents and also automated reasoning
and verification of theories along with the events reconstruction in a
formal model. We argue such an analysis is beneficial to improve the
safety of the passengers and their vehicles, like the airline industry
does for planes.
\end{abstract}

\section{Introduction}

In this work we apply a novel concept of self-forensics
to vehicle design allowing vehicle's subsystems to analyze
themselves forensically as needed and preserve the forensics
data for further automated analysis in the cases of failures,
crash, and event reconstruction.

% We compile a Forensic Lucid specification for self-forensics.

We insist this should be a part of the protocol for
for each new vehicle design, to monitor its hardware and
software components.

\paragraph*{Problem Statement.}

It may become relatively difficult-to-impossible to do
event reconstruction in testing and crash incident analysis
e.g. to determine a faulty component or any other internal cause of crash
if a vehicle in question is badly damaged. Even if some
of information may be available at the incident scene
to the investigators, including eyewitness stories and
the actual final state of the vehicle, it may never be
clear what was the reason (other than the external evidence),
so ad-hoc conclusions can be incorrectly inferred from
what the investigators can see as well as the eyewitness
stories of the exterior to the accident.

\paragraph*{Proposed Solution.}

We propose to include a notion of self-forensic components
included into the hardware and software design of the modern
cars. Their purpose is multi-fold. They can serve as the
real-time monitoring and logging and analyzing subsystems
of critical vehicle parts and overall health as well as
measuring parameters such as as speed, temperature, etc.
over safety-predefined thresholds in a durable blackbox-like
device in a {\flucid} format (with a possibility of external
export via wireless communication) with the intent of
in-situ automatic analysis and reasoning response as well as for after-the-fact
for the purpose of event reconstruction to be complete based
on the self-forensics journals automatically with tools
making sure no event is missed and the analysis is thorough.
The approach, combined with an expert system, can also be
used to train personnel in such investigative techniques.
It would also allow to pin-point a potential cause of
the fault inside rather than external allowing to
more precisely hold someone accountable.

\paragraph*{Organization.}

First, we introduce the notion of forensic computing,
the {\flucid} language, the self-management
properties of autonomous systems, etc. as a background
and the related work to give the reader the idea
for this work in \xs{sect:background-related-work}.
Then we overview the notion of self-forensics
\xs{sect:self-forensics} and its application to
the vehicle safety design with the purpose of
automated reasoning during the activity as well
as preservation of he evidential data for later
analysis and event reconstruction using automated
tools. We go over requirements, limitations,
advantages and examples. Then, we conclude and
present some future work items in \xs{sect:conclusion}.

\section{Background and Related Work}
\label{sect:background-related-work}

\subsection{Forensic Computing}

The notion of {\em computer forensics} also known as {\em cyberforensics}
or the associated {\em forensic computing} has been traditionally associated with
computer crime incidents investigations~\cite{palmer01,imf2008-proceedings}.
We show the approach is useful in vehicle crash investigations and event reconstruction.

Further we argue if the new technologies are built with the self-forensics software
and hardware toolkits or components, it would
even help the automotive industry to improve the safety design of the vehicles and
implement decision-making modules when real-time event analysis may lead
to catastrophic or similar outcomes or during crash testing of new vehicles.

The notion of self-forensics was first proposed
by Mokhov~\cite{self-forensics-flucid-as} for
autonomous systems such as NASA spacecraft as
well as large- and medium-scale distributed
software systems that need to support themselves
and analyze themselves with the new self-forensics
autonomic property.

\subsection{Self-Management Properties}

The notion of self-management and self-managed systems comes
from {\em autonomic computing} (AC)~\cite{autonomic-computing-2006,autonomic-computing-2004}.
The common aspects of self-managing systems, such
as self-healing, self-protection, self-optimization, self-configuration
and the like are now fairly well understood in the literature
and R\&D~\cite{ibmarchblprnt2006,kephartacvis03,truszkowski04,hinchey05,assl-nasa-swarm-sac08}.
We formally introduce another autonomic
property that we call {\em self-forensics} that we would
like to be a part of the standard specification for
vehicle systems specification.

\subsection{{\flucid}}

{\flucid}~\cite{flucid-imf08,flucid-isabelle-techrep-tphols08,marf-into-flucid-cisse08}
is a forensic case specification programming language for automatic deduction
and event reconstruction of computer crime incidents. The language itself is general
enough to specify any events, their properties, duration, as well as the
context-aware system model. We take out {\flucid} from the cybercrime domain
for its eventual application to aid the vehicle crash incidents investigation
from within
as an example of self-forensic case specification.

{\flucid} is based on the
{\lucid}~\cite{lucid85,lucid95,lucid76,lucid77,eager-translucid-secasa08}
language and its various dialects that allow
natural expression of various phenomena, inherently parallel, and
most importantly, {\em context-aware}, i.e. the notion of context
is specified as a first-class value in
{\lucid}~\cite{gipsy-simple-context-calculus-08,wanphd06,tongxinmcthesis08}.
{\lucid} dialects are functional programming languages.
All these properties make {\flucid} an interesting choice
for forensic contextual logging and computing in crash investigations
related to the internal failures of the components.

{\flucid} is also significantly influenced by and is meant
to be a usable improvement of the work of Gladyshev et al. on
formal forensic analysis and event reconstruction using
finite state automate (FSA) to model incidents and reason
about them~\cite{blackmail-case,printer-case}.

While {\flucid} itself is still being finalized as a part of
the PhD work of the author along with its compiler,
run-time and development environments, and the accompanying
expert system, it is well under way
to validate its applicability to various use-cases and
scenarios~\cite{marf-into-flucid-cisse08,self-forensics-through-case-studies}.

\subsubsection*{Context}

{\flucid} is context-oriented.
The basic context entities comprise an observation $o$
in \xe{eq:o}, observation sequence $os$ in \xe{eq:os},
and the evidential statement in \xe{eq:es}. These
terms are inherited from~\cite{blackmail-case,printer-case}
and represent the context of evaluation in {\flucid}.
An observation of a property $P$ has a duration between
$[\min,\min+\max]$ (where $\min$ is the minimum duration
of the observation, $\max$ is a potential variation from
that minimum). This was the original definition of $o$~\cite{blackmail-case,printer-case}
and the author later added $w$ to amend each observation
with weight factor or probability or credibility to
later further model in accordance with the mathematical
theory of evidence~\cite{shafer-evidence-theory}.
$t$ is an optional timestamp as in a forensic log for that property.
An observation sequence represents, which is
a chronologically ordered collection of observations
represent a story witnessed by someone or something
or encodes a description of some evidence. All these
stories (observation sequences, or forensics logs, if you will)
all together represent an evidential statement about
an incident.
The evidential statement is an unordered
collection of observation sequences. The property
$P$ itself can encode anything of interest~--~an element
of any data type or even another {\flucid} expression,
or an object instance hierarchy or an event. It can
be arbitrary ``deep'' in what it can contain as a set
of observer computations, state transitions, complex
objects, outcomes, measurements, etc.

\begin{equation}
o = (P,t,\min,\max,w)
\label{eq:o}
\end{equation}

\begin{equation}
os = \{o_1,\ldots,o_n\}
\label{eq:os}
\end{equation}

\begin{equation}
es = \{os_1,\ldots,os_m\}
\label{eq:es}
\end{equation}

Having constructed the context, one needs to built
a {\em transition function} {\trans} and its inverse {\invtrans}.
The generic versions of them are provided by {\flucid}~\cite{flucid-isabelle-techrep-tphols08}
based on~\cite{printer-case,blackmail-case}, but the investigation-specific
one has to be built, potentially visually (e.g. using a data-flow graph programming tool
that can translate into {\lucid} and back~\cite{yimin04}), by the engineering
team, which can be done even before the vehicle starts,
if the self-forensics aspect is included into the design
from the start. The specific {\invtrans} takes evidential statement
as an argument and the generic {\invtrans} takes the specific {\invtrans}
as an argument as in functional programming.

When the run-time system, such as the General Intensional Programming System
({\gipsy})~\cite{agipsy-ease08,gipsy2005,gipsy-multi-tier-secasa09,gipsy-hoil},
evaluates a {\flucid} specification, navigating
the evidential statement context of evaluation
using intensional operators~\cite{flucid-isabelle-techrep-tphols08},
it traces the execution of the {\invtrans} and computes
the possible backtraces of the events from the final
observed state of the incident back to the known initial
state of the vehicle. The forensic evidence collected
as a context for {\invtrans} is compared against
a potential hypothesis statement, or a witness (can be
device or a sensor or an eyewitness) account to see
if they agree, and if they do, what are the possible
explanations in the backtraces. The backtraces, if exist, are
ordered from the most credible to the least credible
(after computing the aggregated credibility score).
The investigator is then presented with backtraces of
the processed evidence (that can be a large bulk
to sift through manually) representing the reconstructed
events. If there are no such backtraces, then the
witness claim or say manufacturer specification of
a faulty component, do not agree with the evidence,
which may reasonably provably mean that component
is is the cause of the incident or that eyewitness
is lying or the component specification is not what it claims
to be. It is also possible that we do not have enough
evidence in our knowledge base that we acquired from
the incident -- in this case the investigator would
normally know where to look for more evidence or
question more witnesses, etc.

\section{Self-Forensics Requirements for Vehicle Design and Incident Investigation}
\label{sect:self-forensics}

In this section we elaborate in detail on the application
of self-forensics and its requirements that must be made formal,
if the property to be used in the industry.

Existing self-diagnostics, computer BIOS reports,
and {\smart}~\cite{wiki:SMART} reporting for hard disk as well as many other
devices could be a good source for such forensic data computing, i.e. be more forensics-friendly
and provide forensics interfaces for self-forensics analysis and investigation
as well as allowing engineering teams extracting,
analyzing, and reconstructing events using such data.
The process has to be supported by the related languages,
and tools available to the investigator to model the case,
import the actual forensics data gathered, and validate
claims and hypotheses of what happened against, potentially
large and vast volumes of data, potentially automatically
preprocessed with a backtrace of the event reconstruction.

This would be even a greater enhancement and help with the blackboxes,
like in planes in the airline industry, or reasoning about events, possibly speeding
up the analysis of the anomalies in subsystems.

Most, if not all new cars and other automotive road vehicles these days
all have on-board computers and often GPS devices.
This by default reduces the cost of adding of the self-forensics units
to the design and development of hardware and software components
of such vehicles.

\subsection{Application}

The self-forensics property is meant to embody and formalize
all existing and future aspects of self-analysis, self-diagnostics,
data collection and storage, and real-time automatic decision making if necessary that were
not formalized and categorized as such before and define a well-established
category in the industry and academia.
In that view, self-forensics encompasses self-diagnostics, blackbox recording,
(Self-Monitoring, Analysis, and Reporting Technology) {\smart} reporting~\cite{wiki:SMART}, and encoding
this information in analyzable form of {\flucid} contextual logging
for in situ or later automated analysis and
event reconstruction using the corresponding software tool or tools.

Optional parallel logging of the forensics events during
the normal operation of the road vehicles, especially during
``extreme'' periods of operation, i.e. when it is detected that
the speed, internal fluid pressure in any fluid lines, temperature, etc. are
over some minimal soft safety threshold, the granularity and
frequency of reporting may increase can go off-vehicle via
a wireless link to a cell tower or via peer-to-peer
ad-hoc mobile wireless network of vehicles~\cite{performance-mobile-adhoc-net} to eventually be stored
say at the same company that provides the GPS or cellular service
to this particular vehicle or its driver. Shipping such forensics
logs off-site will have a duplicate copy in case the
original gets damaged in the incident. It can also be
used by a more powerful computer at the company to do
a near real-time analysis of the vehicle's performance
and alert the driver if anything anomalous is detected
and predicted to be potentially harmful.

To achieve that all electric, electronic, and mechanical subsystems of a vehicle
can have functional units to observe them other
for anomalies and log them appropriately
for forensics purposes.

\subsection{Training}

Such forensic specification is also useful to train new engineers
on a safety design and testing team,
and others involved, in data analysis, to avoid
potentially overlooking data and making incorrect ad-hoc
decisions.

There are well known log search engines or real-time vulnerability scanners
and detectors that, if adapted to our case, would allow the investigator
searching and even analyzing the multiple logs or even real-time situation,
and even co-relate some events, based on the timestamps, etc.
e.g. such as Splunk~\cite{splunk} and Nessus~\cite{nessus}, adapted
for the use in a vehicle's on-board computer, yet if applied to our case they would
still be tedious and time consuming to work with by an
investigator when examining the logs and detected evidence,
that could be very numerous.

In an {\flucid}-based expert system (that's what
was the original purpose of {\flucid} in the first place
in cybercrime investigations) one can accumulate a number
of contextual facts from the self-forensic evidence and
the trainees can construct their theories of what happened
and see of their theories agree with the evidential data.
Over time, (unlike in most cybercrime investigates) it can accumulate the
general enough contextual knowledge base of encoded facts
that can be analyzed
globally and on the web.
The data can be shared across manufactures
and mechanical and software engineers involved.

\subsection{Requirements}

Here we briefly define the requirements scope for the
self-forensics property adapted to road vehicle design:

\begin{itemize}
\item
Should be optional if constrained by the budget,
but should be strongly recommended to be included.
Must not be optional for mission critical and
safety-critical road vehicles (e.g. military,
police, ambulance, fire trucks, etc.).

\item
Must be included in the design at all times.

\item
Must cover all the self-diagnostics events mentioned earlier
as well as any design-specific events and monitoring.

\item
Must have a formal specification (that what it makes it different from just self-diagnostics).

\item
Must have tools for automated reasoning and reporting about incident analysis
matching the specification.

\item
Context should be synthesized in the terms of system
specification involving the incidents, e.g. parts
and the software and hardware engineering design specification
should be formally encoded (e.g. in {\flucid}).

\item
Preservation of forensic evidence must be atomic, reliable,
robust, and durable.

\item
The forensic data must be able to include any or all related
non-forensic data for analysis when needed,
including measurements taken
around the time of incident
or even the entire trace of a lifetime of a system component
logged somewhere for automated analysis and event
reconstruction.

\item
Levels of forensic logging and detail should be
optionally configurable in collaboration with
other design requirements in order not to hog
other activities, create significant overhead,
or fill in the bandwidth of the wireless connection
or log storage.

\item
Event co-relation optionally should be specified.

\item
Some forensic analysis can be automatically
done by the
vehicle's computer
itself (provided having
enough resources to do so).
\end{itemize}

\subsection{Limitations}
\label{sect:limitations}

The self-forensics autonomic property is
very beneficial to have for automated analysis of incidents in
road vehicles and the like,
but it probably can not be mandated as absolutely required
due to a number of limitations it creates. However, whenever
the monetary and time budgets allow, it should be included
in the design and development of
the road vehicle parts
or software systems capable of self-monitoring and reporting
of encoded forensic data.

Here are some most prominent limitations:

\begin{itemize}
\item
The cost of the vehicle will obviously increase
from manufacturing, to maintenance, and the end-user
buying price.

\item
If built into software, the design and development
requires functional long-term storage and CPU power.

\item
Likely increase of bandwidth requirements; if the
more than twice bandwidth and storage used.

\item
An overhead overall if collect forensics data
continuously. Can be offloaded along with the
control data.

\item
Ideally should be in ROM or similar flash type of
memory; but should allow firmware and software
upgrades.

\item
Current computer logging within vehicles are not
designed to be for the work presented here -- {\flucid}-based
self-forensics analysis, so there will have to be an effort
to adapt it if {\flucid} specifications become an industry standard.
Fortunately, traditional existing logging can be converted
to use the {\flucid} expressions without much
difficulty, that would greatly simplify forensic
analysis of such data.

\item
We do not tackle other autonomic requirements of
the system assuming their complete coverage and
presence in the system from the earlier developments
and literature, such as self-healing, self-protection,
etc.

\item
Transition function {\trans} and its inverse {\invtrans}
has to be modeled by the
engineering team throughout the design phase
and encoded in {\flucid}. Luckily the DFG
IDE~\cite{yimin04}, like a CAD application is to be available.

\item
Manufacturers or some external certifying
agency has to be involved to make sure
the self-forensic components function
as intended to prevent the manufacturer
to circumvent self-incriminating evidence
thereby promoting the quality and
accountability of vehicle manufactures.

\end{itemize}

\subsection{Advantages}

Having {\flucid} helps scripting the forensics events in a log
in the road vehicle's computer or a blackbox.
The blackbox can contain the forensic data encoded anyhow
including forensics expressions, XML, or just compressed
binary and using external tool to convert it to a {\flucid}
specification for further evaluation by investigators.
{\flucid} is context-aware, built upon the intensional logic~\cite{kripke59,kripke69}
and dialect that existed in the literature and math for more then
30 years.

\subsection{Small Example}

\begin{itemize}
\item
	Self-forensic hardware sensors observe
	systems (electrical and mechanic) of a road vehicle.

\item
	Every engineering event is forensically logged.

\item
 	Each forensic sensor may observes several systems or subsystems.

\item
	Each sensor composes a ``story'' of an observational
	sequence a particular system or a component in as a {\flucid} context
	and the component specification is known at the manufacture time
	and encoded as a {\flucid} specification in advance.

\item
	A collection of sensor witness stories from multiple sensors,
	properly encoded, represent the evidential statement.

\item
	An incident happens; engineers define theories about what happened.
	The theories are encoded as observation sequences. When
	evaluated against the evidential statements from all sensors
	the events are co-related and reconstructed according to the {\flucid}
	semantics.

	Then the evaluating system (e.g. {\gipsy})
	can automatically verify the theory automatically against
	the context of evidential statement and if the theory $T$ agrees
	with the evidence, meaning this theory has an explanation
	within the given evidence (and the amount of evidence can
	be significantly large for ``eyeballing'' it by humans),
	then likely the theory is a possible explanation of what
	has happened. It is possible to have multiple explanations
	and multiple theories agreeing with the evidence. In the
	latter case usually the ``longer'' (in the amount of
	events and observations involved) theory is preferred
	or the one that has a higher cumulative weight/credibility
	$w$.
\end{itemize}

% \section{Example {\flucid} Specification of Self-Forensics}
%
% {\todo}
%
%
% \begin{verbatim}
% invtrans(T @ miata_es_xyz123)
% where
%   evidential statement miata_es_xyz123 = {...};
%
%   // T is a theory of what
%   observation sequence T = {};
% end
% \end{verbatim}

\section{Conclusion and Future Work}
\label{sect:conclusion}

We would like to conclude with the remarks that we stated
at the beginning -- that the self-forensics modules and
components, both hardware sensors, a simpler analogue of a blackbox,
and software components
should be included into the modern road vehicle design
for real-time and post-mortem forensic log data analysis
in the {\flucid}-encoded form to aid investigation and
event reconstruction with the formal approach to make
the investigation complete not only from the outside, but
also from the inside by using the forensics data.
Furthermore, such data can be used during the crash testing
of the road vehicles by the manufacturer as well as to
train the engineering teams and the investigators to do
a more complete analysis with an aid of the {\flucid}-based
expert system.

We also discuss the potential limitations of the approach.
In our future work we will attempt to address the limitations
as well as complete some other intermediate items along the
way, specifically.

%\section{Future Work}
\paragraph*{Future Work.}

\begin{itemize}
\item
Estimate the feasibility of ``back-porting'' the self-forensics
modules and components into the existing fleets of road vehicles
as well as the costs of storage and maintenance of such data,
when it expires, and who has a control over it.

\item
Amend the Autonomic System Specification Language ({\assl})~\cite{assl-sew,assl-seams,vassevicsoft06,vassevPhDThesis}
to handle the self-forensic property along with its formal
specification and verification.

\item
Implement the notion of self-forensics in the
{\gipsy}~\cite{agipsy-ease08,gipsy2005,gipsy-multi-tier-secasa09,gipsy-hoil,assl-generated-arch-for-as}
and
{\dmarf}~\cite{dmarf-web-services-cisse08,dmarf-assl-self-healing,dmarf-assl-self-protection,dmarf-assl-self-optimization}
systems the author is closely working on.

\item
Finalize implementation of the {\flucid} compiler
and the development and run-time environment.

\item
Implement large realistic cases encoded in {\flucid}
to test and validate various aspects of correctness,
performance, and usability.

\item
Industrialize and standardize the concept in the
industry.

\item
Resolve and map out privacy issues with external
reporting of the forensic evidence via the wireless
transmission and the applicable laws.

\item
Durability.

\item
Cost analysis of the self-forensic system deployment
in a road vehicle.

\item
This research can aid crash investigations not only in
automotive vehicles, but technically in all vehicle
types, e.g. in the aviation industry by incorporating
the self-forensics features and components into
helicopters, planes, etc. to aid crash
investigations such as~\cite{california-bus-crash-apr-2009,scotland-helicopter-crash-failure}.

\end{itemize}

% References
\bibliographystyle{unsrt}
\bibliography{self-forensics-with-flucid-cmrsc}

\begin{thebibliography}{10}

\bibitem{palmer01}
G.~Palmer (Editor).
\newblock A road map for digital forensic research, report from first digital
  forensic research workshop ({DFRWS}).
\newblock Technical report, {DFRWS}, 2001.

\bibitem{imf2008-proceedings}
Oliver G{\"o}bel, Sandra Frings, Detlef G{\"u}nther, Jens Nedon, and Dirk
  Schadt, editors.
\newblock {\em IT-Incidents Management {\&} IT-Forensics - IMF 2008, Conference
  Proceedings, September 23-25, 2008, Mannheim, Germany}, volume 140 of {\em
  LNI}. GI, 2008.

\bibitem{self-forensics-flucid-as}
Serguei~A. Mokhov et~al.
\newblock Self-forensics for autonomous systems.
\newblock Submitted for publication at IEEE Com. Mag., 2009.

\bibitem{autonomic-computing-2006}
M.~Parashar and S.~Hariri, editors.
\newblock {\em Autonomic Computing: Concepts, Infrastructure and Applications}.
\newblock CRC Press, December 2006.

\bibitem{autonomic-computing-2004}
R.~Murch.
\newblock {\em Autonomic Computing: On Demand Series}.
\newblock IBM Press, Prentice Hall, 2004.

\bibitem{ibmarchblprnt2006}
{IBM Corporation}.
\newblock An architectural blueprint for autonomic computing.
\newblock Technical report, {IBM Corporation}, 2006.

\bibitem{kephartacvis03}
Jeffrey~O. Kephart and David~M. Chess.
\newblock The vision of autonomic computing.
\newblock {\em IEEE Computer}, 36(1):41--50, 2003.

\bibitem{truszkowski04}
Walt Truszkowski, Mike Hinchey, James Rash, and Christopher Rouff.
\newblock {NASA}'s swarm missions: The challenge of building autonomous
  software.
\newblock {\em IT Professional}, 6(5):47--52, 2004.

\bibitem{hinchey05}
Michael~G. Hinchey, James~L. Rash, Walter Truszkowski, Christopher Rouff, and
  Roy Sterritt.
\newblock Autonomous and autonomic swarms.
\newblock In {\em Software Engineering Research and Practice}, pages 36--44.
  CSREA Press, 2005.

\bibitem{assl-nasa-swarm-sac08}
Emil Vassev, Michael~G. Hinchey, and Joey Paquet.
\newblock Towards an {ASSL} specification model for {NASA} swarm-based
  exploration missions.
\newblock In {\em Proceedings of the 23rd Annual ACM Symposium on Applied
  Computing (SAC 2008) - AC Track}. ACM, 2008.

\bibitem{flucid-imf08}
Serguei~A. Mokhov, Joey Paquet, and Mourad Debbabi.
\newblock Formally specifying operational semantics and language constructs of
  {Forensic Lucid}.
\newblock In Oliver G{\"o}bel, Sandra Frings, Detlef G{\"u}nther, Jens Nedon,
  and Dirk Schadt, editors, {\em Proceedings of the IT Incident Management and
  IT Forensics (IMF'08)}, pages 197--216, Mannheim, Germany, September 2008.
  GI.
\newblock {LNI140}.

\bibitem{flucid-isabelle-techrep-tphols08}
Serguei~A. Mokhov and Joey Paquet.
\newblock Formally specifying and proving operational aspects of {Forensic
  Lucid} in {Isabelle}.
\newblock Technical Report 2008-1-Ait Mohamed, Department of Electrical and
  Computer Engineering, Concordia University, August 2008.
\newblock In Theorem Proving in Higher Order Logics (TPHOLs2008): Emerging
  Trends Proceedings.

\bibitem{marf-into-flucid-cisse08}
Serguei~A. Mokhov.
\newblock Encoding forensic multimedia evidence from {MARF} applications as
  {Forensic Lucid} expressions.
\newblock In {\em Proceedings of {CISSE'08}}, University of Bridgeport, CT,
  USA, December 2008. Springer.
\newblock To appear.

\bibitem{lucid85}
William Wadge and Edward Ashcroft.
\newblock {\em Lucid, the Dataflow Programming Language}.
\newblock Academic Press, London, 1985.

\bibitem{lucid95}
Edward Ashcroft, Anthony Faustini, Raganswamy Jagannathan, and William Wadge.
\newblock {\em Multidimensional, Declarative Programming}.
\newblock Oxford University Press, London, 1995.

\bibitem{lucid76}
Edward~A. Ashcroft and William~W. Wadge.
\newblock Lucid - a formal system for writing and proving programs.
\newblock {\em SIAM J. Comput.}, 5(3), 1976.

\bibitem{lucid77}
Edward~A. Ashcroft and William~W. Wadge.
\newblock Erratum: {Lucid} - a formal system for writing and proving programs.
\newblock {\em SIAM J. Comput.}, 6((1):200), 1977.

\bibitem{eager-translucid-secasa08}
John Plaice, Blanca Mancilla, Gabriel Ditu, and William~W. Wadge.
\newblock Sequential demand-driven evaluation of eager {TransLucid}.
\newblock In {\em Proceedings of the 32nd Annual {IEEE} International Computer
  Software and Applications Conference ({COMPSAC})}, pages 1266--1271, Turku,
  Finland, July 2008. IEEE Computer Society.

\bibitem{gipsy-simple-context-calculus-08}
Joey Paquet, Serguei~A. Mokhov, and Xin Tong.
\newblock Design and implementation of context calculus in the {GIPSY}
  environment.
\newblock In {\em Proceedings of the 32nd Annual {IEEE} International Computer
  Software and Applications Conference ({COMPSAC})}, pages 1278--1283, Turku,
  Finland, July 2008. IEEE Computer Society.

\bibitem{wanphd06}
Kaiyu Wan.
\newblock {\em Lucx: {Lucid} Enriched with Context}.
\newblock PhD thesis, Department of Computer Science and Software Engineering,
  Concordia University, Montreal, Canada, 2006.

\bibitem{tongxinmcthesis08}
Xin Tong.
\newblock Design and implementation of context calculus in the {GIPSY}.
\newblock Master's thesis, Department of Computer Science and Software
  Engineering, Concordia University, Montreal, Canada, April 2008.

\bibitem{blackmail-case}
Pavel Gladyshev.
\newblock Finite state machine analysis of a blackmail investigation.
\newblock {\em International Journal of Digital Evidence}, 4(1), 2005.

\bibitem{printer-case}
Pavel Gladyshev and Ahmed Patel.
\newblock Finite state machine approach to digital event reconstruction.
\newblock {\em Digital Investigation Journal}, 2(1), 2004.

\bibitem{self-forensics-through-case-studies}
Serguei~A. Mokhov et~al.
\newblock Self-forensics through case studies of small to medium software
  systems.
\newblock Unpublished, 2009.

\bibitem{shafer-evidence-theory}
G.~Shafer.
\newblock {\em The Mathematical Theory of Evidence}.
\newblock Princeton University Press, 1976.

\bibitem{yimin04}
Yi~Min Ding.
\newblock Bi-directional translation between data-flow graphs and {Lucid}
  programs in the {GIPSY} environment.
\newblock Master's thesis, Department of Computer Science and Software
  Engineering, Concordia University, Montreal, Canada, 2004.

\bibitem{agipsy-ease08}
Emil Vassev and Joey Paquet.
\newblock Towards autonomic {GIPSY}.
\newblock In {\em Proceedings of the Fifth IEEE Workshop on Engineering of
  Autonomic and Autonomous Systems ({EASE} 2008)}, pages 25--34, Los Alamitos,
  CA, USA, 2008. IEEE Computer Society.

\bibitem{gipsy2005}
Joey Paquet and Ai~Hua Wu.
\newblock {GIPSY} -- a platform for the investigation on intensional
  programming languages.
\newblock In {\em Proceedings of the 2005 International Conference on
  Programming Languages and Compilers ({PLC} 2005)}, pages 8--14, Las Vegas,
  USA, June 2005. CSREA Press.

\bibitem{gipsy-multi-tier-secasa09}
Joey Paquet.
\newblock A multi-tier architecture for the distributed eductive execution of
  hybrid intensional programs.
\newblock In {\em Proceedings of 2nd IEEE Workshop in Software Engineering of
  Context Aware Systems ({SECASA}'09)}. IEEE Computer Society, 2009.
\newblock To appear.

\bibitem{gipsy-hoil}
Serguei~A. Mokhov and Joey Paquet.
\newblock Using the {General Intensional Programming System (GIPSY)} for
  evaluation of higher-order intensional logic ({HOIL}) expressions.
\newblock Submitted for publication at LFTMP'09, 2009.

\bibitem{wiki:SMART}
Wikipedia.
\newblock {S.M.A.R.T.} --- {Wikipedia}{,} the free encyclopedia.
\newblock [Online; accessed 9-February-2009], 2009.
\newblock
  \url{http://en.wikipedia.org/w/index.php?title=S.M.A.R.T.&oldid=269322389}.

\bibitem{performance-mobile-adhoc-net}
Tatsuaki Osafune, Monden Kazuya, Shoji Fukuzawa, and Susumu Matsui.
\newblock Performance measurement of mobile ad hoc network for application to
  internet-its (intelligent transportation system).
\newblock In {\em SAINT}, pages 25--30, 2004.

\bibitem{splunk}
{Splunk Inc.}
\newblock Splunk: Search and analysis engine for {IT} data.
\newblock [online], 2005--2009.
\newblock \url{http://www.splunk.com/}.

\bibitem{nessus}
{Tenable Network Security}.
\newblock Nessus: the network vulnerability scanner.
\newblock [online], 2002--2009.
\newblock \url{http://www.nessus.org/nessus/}.

\bibitem{kripke59}
Saul~A. Kripke.
\newblock A completeness theorem in modal logic.
\newblock {\em Journal of Symbolic Logic}, 31(2):276--277, 1966.

\bibitem{kripke69}
Saul~A. Kripke.
\newblock Semantical considerations on modal logic.
\newblock {\em Journal of Symbolic Logic}, 34(3):501, 1969.

\bibitem{assl-sew}
Emil Vassev and Joey Paquet.
\newblock {ASSL -- Autonomic System Specification Language}.
\newblock In {\em Proceedings if the 31st Annual {IEEE} / {NASA} Software
  Engineering Workshop ({SEW-31})}, pages 300--309, Baltimore, MD, USA, March
  2007. {NASA/IEEE}.

\bibitem{assl-seams}
Emil Vassev and Joey Paquet.
\newblock Towards an autonomic element architecture for {ASSL}.
\newblock In {\em Proceedings of the 29th {IEEE} International Conference on
  Software Engineering / Software Engineering for Adaptive and Self-managing
  Systems ({ICSE 2007 SEAMS})}, page~4, Minneapolis, MN, USA, May 2007. {IEEE}.

\bibitem{vassevicsoft06}
Emil Vassev, Heng Kuang, Olga Ormandjieva, and Joey Paquet.
\newblock Reactive, distributed and autonomic computing aspects of {AS-TRM}.
\newblock In {\em Proceedings of the 1st International Conference on Software
  and Data Technologies - {ICSOFT}'06}, pages 196--202, 2006.

\bibitem{vassevPhDThesis}
Emil~I. Vassev.
\newblock {\em Towards a Framework for Specification and Code Generation of
  Autonomic Systems}.
\newblock PhD thesis, Department of Computer Science and Software Engineering,
  Concordia University, Montreal, Canada, 2008.

\bibitem{assl-generated-arch-for-as}
Emil Vassev and Serguei~A. Mokhov.
\newblock An {ASSL}-generated architecture for autonomic systems.
\newblock In {\em Proceedings of C3S2E'09}. ACM, 2009.
\newblock To appear.

\bibitem{dmarf-web-services-cisse08}
Serguei~A. Mokhov and Rajagopalan Jayakumar.
\newblock Distributed modular audio recognition framework ({DMARF}) and its
  applications over web services.
\newblock In {\em Proceedings of TeNe'08}. Springer, 2008.
\newblock To appear.

\bibitem{dmarf-assl-self-healing}
Emil Vassev and Serguei~A. Mokhov.
\newblock Towards autonomic specification of {Distributed MARF} with {ASSL}:
  Self-healing.
\newblock Submitted for publication to Middleware'09, 2009.

\bibitem{dmarf-assl-self-protection}
Serguei~A. Mokhov and Emil Vassev.
\newblock Autonomic specification of self-protection for {Distributed MARF}
  with {ASSL}.
\newblock In {\em Proceedings of C3S2E'09}. ACM, 2009.
\newblock To appear.

\bibitem{dmarf-assl-self-optimization}
Emil Vassev and Serguei~A. Mokhov.
\newblock Self-optimization property in autonomic specification of {Distributed
  MARF} with {ASSL}.
\newblock Submitted for publication to ICSOFT'09, 2009.

\bibitem{california-bus-crash-apr-2009}
{CNN}.
\newblock California bus crash.
\newblock [online], April 2009.
\newblock \url{http://www.cnn.com/2009/US/04/29/california.crash/index.html}.

\bibitem{scotland-helicopter-crash-failure}
{CNN}.
\newblock `{Catastrophic} failure' caused {North Sea} copter crash.
\newblock [online], April 2009.
\newblock
  \url{http://www.cnn.com/2009/WORLD/europe/04/11/scotland.helicopter.crash.fa%
ilure/index.html}.

\end{thebibliography}

\end{document}